\documentclass[12pt,a4paper]{article}
\usepackage[margin=2cm]{geometry}
\usepackage[utf8]{inputenc}
\usepackage[T2A]{fontenc}
\usepackage[english]{babel}
\usepackage{amssymb}
\usepackage{amsmath,mathtools}
\usepackage{authblk}
\usepackage{mathrsfs}
\usepackage{graphicx}
\usepackage{cite}
\usepackage{indentfirst}
\usepackage{color}
\usepackage[colorlinks=true,linkcolor={blue},citecolor={green},urlcolor={red}]{hyperref}

\usepackage{caption2}[2008/03/29]    

\newcommand{\mathdash}{\,\text{---}\,}

\newcommand{\diff}[2]{{\frac{d{#1}}{d{#2}}}}
\newcommand{\pdiff}[2]{{\frac{\partial{#1}}{\partial{#2}}}}

\begin{document}

\renewcommand{\figurename}{Fig.}
\renewcommand{\tablename}{Table}
\renewcommand{\captionlabeldelim}{.~}

\title{On the possible electromagnetic manifestations of merging black holes}

\author[1]{D. V. Bisikalo\thanks{bisikalo@inasan.ru}}
\author[1]{A. G. Zhilkin\thanks{zhilkin@inasan.ru}}
\author[1]{E. P. Kurbatov\thanks{kurbatov@inasan.ru}}
\affil[1]{Institute of Astronomy, RAS, Moscow, Russia}
\date{}

\maketitle
\begin{abstract}
We consider scenario of merger of two stellar mass black holes surrounded by an accretion disk. Due to emission of gravitational waves, the mass of the central object decreases and accretion disk experiences perturbation. Calculations show that the main consequence of this disturbance is formation of a shock wave propagating from the center of the disk to its periphery. Light curve is computed and duration of the flash is estimated under assumption that the flash terminates when the luminosity returns to the initial value.

It is shown that, if the total mass of the merging binary is $55\,M_\odot$ (like in the event GW170814), the flash produced by the shock will increase bolometric luminosity of the disk by $4\mathdash6$ orders of magnitude, up to $10^{45}$erg$/$s (absolute stellar magnitude $-23.8^\mathrm{m}$). With account of the distance to the source ($540$~Mpc) and for reasonable assumptions on the parameters of the accretion disk, it turns out that the apparent magnitude of the flash at the maximum of the spectral flux density should be $12.8^\mathrm{m} \mathdash 14.2^\mathrm{m}$, while  duration of the flash --- few minutes.

The main part of the shock radiation flux is emitted in the X-ray and gamma-ray ranges. In the spectral band of the EPIC instrument of the XMM-Newton observatory or the telescope eROSITA of the Spectrum-RG observatory ($ 0.3\mathdash 10$ keV), luminosity will increase by $3 \mathdash 4$ orders of magnitude ($7.5^\mathrm{m} \mathdash 10^\mathrm{m}$), up to $10^{44}$~erg$/$s, corresponding to the apparent stellar magnitude about $17^\mathrm {m}$. Luminosity is at maximum in the observational band of the IBIS instrument of the INTEGRAL observatory ($20\text{~keV} \mathdash 10$ MeV) and will be $10^{44}\mathdash 10^{45}$~erg$/$s, corresponding to the apparent flux $10^{-4}$~photons per $/$cm$^2/$s$/$keV  at the wavelength $\sim 100$~keV. From the far UV  to the longer wavelengths, the brightening is virtually absent --- at the wavelength $10$~eV luminosity, approximately, doubles and corresponds to the absolute magnitude  $-6^\mathrm{m}$ and visual one $32^\mathrm{m}$.
\end{abstract}

\section{Introduction}

By the time of writing (July 2018), gravitational waves observatory LIGO detected 5 mergers of stellar mass black holes. The mergers were accompanied by  emission of $2.3\%\mathdash5.8\%$ of the rest energy of the original binary system as gravitational waves. The parameters of merging objects are listed in Table~\ref{tbl-sources} (masses of objects, fraction of lost mass, distance) \cite{Abbott2016ApJ...818L..22A,Abbott2016PhRvL.116x1103A,Abbott2017PhRvL.118v1101A,Abbott2017ApJ...851L..35A,Abbott2017PhRvL.119n1101A}.

\begin{table}
  \centering
  \begin{tabular}{lccc}
    \hline
    Object & $M_1$, $M_2$ [$M_\odot$] & $\Delta M/(M_1 + M_2)$ [\%] & $D$ [Mpc]  \\
    \hline \\[-10pt]
    GW150914 \cite{Abbott2016ApJ...818L..22A}
      & $36_{-4}^{+5}$, $29_{-4}^{+4}$ & $3.3 \mathdash 5.2$ & $410_{-180}^{+160}$  \\[5pt]
    GW151226 \cite{Abbott2016PhRvL.116x1103A}
      & $14.2_{-3.7}^{+8.3}$, $7.5_{-2.3}^{+2.3}$ & $2.9 \mathdash 5.5$ & $440_{-190}^{+180}$  \\[5pt]
    GW170104 \cite{Abbott2017PhRvL.118v1101A}
      & $31.2_{-6.0}^{+8.4}$, $19.4_{-5.9}^{+8.3}$ & $2.3 \mathdash 5.7$ & $880_{-390}^{+450}$  \\[5pt]
    GW170608 \cite{Abbott2017ApJ...851L..35A}
      & $12_{-2}^{+7}$, $7_{-2}^{+2}$ & $2.8 \mathdash 5.1$ & $340_{-140}^{+140}$  \\[5pt]
    GW170814 \cite{Abbott2017PhRvL.119n1101A}
      & $30.5_{-3.0}^{+5.7}$, $25.3_{-4.2}^{+2.8}$ & $4.0 \mathdash 5.8$ & $540_{-210}^{+130}$  \\[5pt]
  \end{tabular}
  \caption{The parameters of merging binary BHs according to the data of gravitational waves observations. In the columns are listed the names of sources, masses of components, relative variation of the mass of the source, distance to the source.}
  \label{tbl-sources}
\end{table}

In the studies of black holes (BH), it is often assumed that they are surrounded by accretion disks. As the source of the disk matter can serve molecular cloud through which BH flies, the wind from neighboring astrophysical objects, the remnants of a star destroyed by a tide, etc. Existence of an accretion disk around a binary BH is quite possible too. Obviously, such a disk could not form when the stars were on the main-sequence or were giants --- BH precursors, since the strong stellar wind and supernovae explosions would destroy the disk. Formation of an accretion disk at the binary BH stage seems to be possible, if the lifetime of the system before merger was large enough; in this case, the same mechanisms that result in formation of a disk around a single BH can be efficient. The lifetime of a binary BH to the confluence can be estimated as the characteristic time of loss of the angular momentum by a close binary system with circular orbits
\cite{Tutukov2017ARep...61..833T}:
\begin{equation}
  \label{eq-tau-gwr}
  \tau_\mathrm{GWR}
   \approx 10^8 \left( \frac{a}{R_\odot} \right)^4
    \frac{M_\odot^3}{M_1 M_2\,(M_1 + M_2)}~\text{yr},
\end{equation}
where $a$ is separation of components,  $M_1$ and $M_2$ --- masses of components. In order for a binary BH with the parameters similar to GW170814 \cite{Abbott2017PhRvL.119n1101A} (the latest detected event to date) to merge in the Hubble time 13.7\,Gyr, it is necessary to have initial separation less than $48\,R_\odot$. Thus, under any reasonable assumption about the initial distance between the components of a binary BH, the system lifetime will be sufficient to form a circumbinary accretion disk.

The differences of the accretion disks around binary stars and disks around single stars are smallish \cite{Kaigorodov2010ARep...54.1078K,Sytov2011ARep...55..793S} and represent, mainly, a rather large internal radius of the disk, of about $1.5 \mathdash 2$ separations of system components. Soaring merger of a binary BH begins when the distance between components is $a \approx 6\,r_\mathrm{g}$, where $r_\mathrm{g}$ is the gravitational radius of the accretor, i.e. the internal radius of the accretion disk $r_\mathrm{in}$ is of the order of $ 10\,r_\mathrm{g} $. This allows to assume that the disk around a stellar mass binary BH is standard (non-relativistic) and its features can be described using the conventional $\alpha$-disk model of Sha\-ku\-ra-Su\-nya\-ev \cite{Shakura1973A&A....24..337S}. Using this model, we can immediately estimate the luminosity of the original disk: under fairly conservative assumptions about its parameters (the temperature of the inner part of the disk is of the order of $10^7$\,K, accretion rate does not exceed several hundredths of the critical value), luminosity in the soft and medium X-ray ranges will be $10^{40 \pm 1}$\,erg$/$s.

It was shown in many theoretical studies that the merger of a binary BH and the subsequent gravitational-wave radiation  directly affects accretion disk in several ways. First, mass loss by the central compact object after merger leads to a violation of (quasi)-Keplerian equilibrium in the disk and subsequent excitation of high amplitude waves \cite{Bode2007APS..APR.S1010B}. Second, depending on the ratio of the masses of the merging objects and orientation of their own angular momenta, gravitational-wave radiation can be asymmetric. As a result, the matter of the disc gets a recoil momentum that can lead to supersonic perturbations \cite{Bekenstein1973ApJ...183..657B}. Third, the changes in the space-time metric in gravitational waves can directly cause mechanical stresses in the disk that dissipate in the viscous time scale \cite{Kocsis2008PhRvL.101d1101K}. All these phenomena lead to an electromagnetic response of the accretion disk: brightening, quasiperiodic variations of luminosity.

A large number of papers on this phenomenon concerned supermassive BHs in the nuclei of merging galaxies (see, for instance, \cite{Megevand2009PhRvD..80b4012M,ONeill2009ApJ...700..859O,Corrales2010MNRAS.404..947C,Rosotti2012MNRAS.425.1958R}). So, if a binary BH with a total mass of $10^6\,M_\odot$ loses $5\%$ of its mass after the merger, luminosity of the accretion disk should increase by an order of magnitude, to $10^{43}$\,erg$/$ \cite{Corrales2010MNRAS.404..947C}. The effect of the recoil leads to a comparable increase in luminosity, but its magnitude depends significantly on the poorly defined parameters of the merging binary \cite{Fitchett1983MNRAS.203.1049F,Pietila1995CeMDA..62..377P}.

As an example of a study of stellar mass BH we can mention the paper by de Mink et al. \cite{DeMink2017ApJ...839L...7D}, who obtained approximate estimates of the luminosity of perturbed disk as a function of the parameters of the disk and the binary. Typical luminosity appeared to be $\lesssim 10^{43}$\,erg$/$s, with a maximum in the X-ray range of electromagnetic spectrum.

The main aim of the paper is to investigate the response of the accretion disk around binary BH to the decrease of BH mass after the merger. To study this effect, we developed a numerical model model and performrd a series of simulations for disks with dominant gas pressure. The analysis of validity of the model was carried out, the spectra of electromagnetic radiation were calculated and the estimates of the duration of the flash were obtained. As it turned out, the bulk of the energy of electromagnetic radiation is emitted in the X-ray and gamma-ray ranges, where the luminosity reaches $10^{46}$\,erg$/$s. Conditions for detection of the electromagnetic signal by the observatories XMM-Newton and INTEGRAL were determined.

The paper is structured as follows. The next section describes the parameters of stationary accretion disks used in this work. In the third section a model for numerical solution of an unsteady problem is described. In the fourth section calculations of the spectrum of electromagnetic radiation are presented and the possibilities of observing of simulated systems are considered.
Conclusions are presented in the fifth section.

\section{Formulation of the problem}
\label{sec-setting}

One of the universal models of stationary disk accretion is the standard Sha\-ku\-ra-Su\-nya\-ev $\alpha$-disk model \cite{Shakura1973A&A....24..337S}. This model describes a geometrically thin disc in which dissipation and angular momentum transfer are provided by turbulent viscosity. Turbulence is characterized by a single parameter $\alpha \leq 1$. The rate of dissipation of angular momentum affects both accretion rate and the rate of energy release in the disk and, ultimately, determines density and temperature distribution across the disk.

The structure of accretion disk in the standard Sha\-ku\-\-ra-Su\-nya\-ev model depends on the following parameters: dimensionless mass of the accretor $m \equiv M/M_\odot$, dimensionless accretion rate $\dot{m} \equiv \dot{M}/\dot{M}_\mathrm{cr}$, efficiency of dissipation of angular momentum (turbulence parameter) $\alpha$, efficiency of the emission of gravitational energy (accretion efficiency) $\eta$, inner radius of the accretion disk $r_\mathrm{in}$.

In the case of binary BH under consideration, before the merger the mass of the accretor is equal to the total mass of the components of the system, $M \equiv M_1 + M_2$. Critical rate of accretion is determined by the Eddington luminosity \cite{Shakura1973A&A....24..337S}:
\begin{equation}
  \frac{\dot{M}_\mathrm{cr}}{M_\odot/\text{yr}}
  = 3\cdot10^{-8}\,\frac{0.06}{\eta}\,m.
\end{equation}
In the case of a single non-rotating black hole the parameter $\eta$ is approximately $0.06 \mathdash 0.08$ \cite{Lipunov1992ans..book.....L}. Account of rotation increases the efficiency of stationary accretion to $\sim 0.32$ \cite{Bardeen1970Natur.226...64B,Thorne1974ApJ...191..507T}. In the scenario of episodic accretion, efficiency can reach $0.43$ \cite{Li2000ApJ...534L.197L}. However, the value of this parameter, is not critical in this model, because a decisive role belongs to the dimensionless rate of accretion $\dot{m}$.

Presumable inner radius of the disk was obtained by numerical simulation of accretion in a particular binary BH  \cite{Bowen2018ApJ...853L..17B}. It follows from these calculations that the inner radius of the disk is, approximately, equal to the doubled separation of components. It is not difficult to show that this estimate is also true for any binary BH. Indeed, gravitational radius of the accretor can be written as
\begin{equation}
  r_\mathrm{g} = \frac{2GM}{c^2}
  \qquad\text{or}\qquad
  \frac{r_\mathrm{g}}{R_\odot} = 4.2\cdot 10^{-6}\,m  \;.
\end{equation}
Binary BH begins to merge when the distance between its components is $a \approx 6\,r_\mathrm{g}$. Close to the inner boundary of the accretion disk, gas flow is disturbed by the tidal force from the binary BH side. Position of $r_\mathrm{in}$ can be estimated from the condition of equality of tidal force and gravitational force from a point source of the  mass $M$:
\begin{equation}
  \frac{2GM}{r_\mathrm{in}^2}\,\frac{a/2}{r_\mathrm{in}}
  = \frac{GM}{r_\mathrm{in}^2}  \;.
\end{equation}
It follows from this expression that $r_\mathrm{in} = 6\,r_\mathrm{g}$ is providing the estimate of the lower bound of the internal radius of the disk, since the gas-dynamic effects lead to an increase of $r_\mathrm{in}$. Below, we will assume $r_\mathrm{in} = 10\,r_\mathrm{g}$. This estimate agrees with the results of Artymowicz \cite{Artymowicz1994ApJ...421..651A} for a binary system with circular orbits of components and mass ratio of components close to $1$. For $m = 55$, we get $r_\mathrm{in} = 1.6 \cdot 10^8$~cm.

Depending on the accretion rate $\dot{m}$ (as well as on $\alpha$ and $m$, but to a lesser degree), one may distinguish in the disk several zones in which either gas or radiation pressure prevails and which differ in the mechanisms of absorption of radiation. In this paper, we restrict ourselves to the consideration of disks in which the gas pressure prevails in the equilibrium state. According to \cite{Shakura1973A&A....24..337S}, these are the so-called ``B'' and ``C'' zones. A condition for the absence in the disk of a region where the radiation pressure is significant (``A'' zone) looks as%
\footnote{%
The factor in the r.h.s. of the expression \eqref{eq-no-zone-a} differs from the corresponding factor in the formula (2.18) in \cite{Shakura1973A&A....24..337S}: first, in the original article $r_\mathrm{in} = 3 r_\mathrm{g}$; second, we have ascertained this multiplier by numerical calculation.}:
\begin{equation}
  \label{eq-no-zone-a}
  \dot{m} < \dot{m}_\mathrm{low} \equiv \frac{7}{170}\,\frac{1}{(\alpha m)^{1/8}}  \;.
\end{equation}
Dependence of $\dot{m}$ on $\alpha m$ is rather weak. Thus, for $m = 55$ and $0.001 \leqslant \alpha \leqslant 1$ it is equal to $0.025 \lesssim \dot{m}_\mathrm{low} \lesssim 0.059$.

Characteristic temperature scale in the zones ``B'' and ``C'' is of the order of $10^6 \mathdash 10^7\,\text{K}$, while  the densities are of the order of $1$\,g$/$cm$^{3}$. Shakura and Sunyaev derived radial profiles of the bulk density of gas averaged over the thickness of the disk, as well as the surface temperature (at the optical depth of the order of unity). The temperature in the inner layers of the disk depends on the efficiency of heat transfer. Under conditions of ``B''- and ``C''-zones, absorption of radiation is determined  both by Thomson scattering and free-free transitions. Their cross-sections are, respectively, $\sigma_\mathrm{T} = 0.40$\,cm$^2/$g and $\sigma_\mathrm{ff} = 0.11 N/T^{7/2}$\,cm$^2/$g, where $N$ is electron number density. Optical thickness of the disk in the vertical direction can be estimated as $\tau = (\sigma_\mathrm{T} + \sigma_\mathrm{ff})\,\Sigma \gtrsim 10^5$, while $\sigma_\mathrm{T} \gg \sigma_\mathrm{ff}$. Under conditions of large optical thickness, surface temperature $T_\mathrm{s}$ (at the optical depth of the order of $1$) and the temperature in the middle of the disk, $T$, are related as \cite{Dong2016ApJ...823..141D}
\begin{equation}
  T^4 \approx \frac{3\tau}{8}\,T_\mathrm{s}^4  \;.
\end{equation}
Thus, if the thermal energy is transferred by radiative thermal conductivity, $T/T_\mathrm{s} \gtrsim 10$.

If convection is responsible for the energy transfer in the vertical direction, temperature difference may be estimated assuming that gas distribution is isentropic:
\begin{equation}
  \frac{T}{T_\mathrm{s}} = \left( \frac{\rho}{\rho_\mathrm{s}} \right)^{\gamma-1}  \;,
\end{equation}
where $\rho$ and $\rho_\mathrm{s}$ are bulk gas density in the middle and at the surface of the disk, respectively;
$\gamma$ --- adiabatic exponent. According to the estimates obtained in \cite{Shakura1973A&A....24..337S}, $\rho/\rho_\mathrm{s} \approx 2$. Then, for monatomic gas ($\gamma = 5/3$), one obtains $T/T_\mathrm{s} \approx 1.6$.

The timescale of radiative thermal conductivity is
\begin{equation}
  t_\mathrm{rad} \sim \tau\,\frac{H}{c} \sim \frac{\tau}{\Omega}\,\frac{c_\mathrm{s}}{c}  \;,
\end{equation}
where $\tau$ --- optical thickness of the disk (see above); $H \sim c_\mathrm{s}/\Omega$ --- semi-thickness of the disk; $c_\mathrm{s}$ --- sound speed; $\Omega$ --- Keplerian frequency. The timescale of convection may be estimated as
\begin{equation}
  t_\mathrm{conv} \sim \frac{H}{\alpha c_\mathrm{s}} \sim \frac{1}{\alpha \Omega}  \;.
\end{equation}
The ratio of timescales shows which process of energy transfer is more efficient:
\begin{equation}
  \frac{t_\mathrm{rad}}{t_\mathrm{conv}}
  \sim \alpha \tau\,\frac{c_\mathrm{s}}{c}
  \sim 400\,\alpha \left( \frac{T}{10^7 \text{~К}} \right)^{1/2}  \;.
\end{equation}
It is seen that in zones ``B'' and ``C'' the energy is transferred more efficiently by convection. Based on the estimates obtained above, we will deem below that the ratio of temperatures in the middle of the disk and at its surface is
\begin{equation}
  \label{eq-midplane-surface-temperature}
  \frac{T}{T_\mathrm{s}} \equiv 2  \;.
\end{equation}

Let set equilibrium profiles of the temperature and density as power functions of the radial coordinate:
\begin{align}
  \label{eq-temp-power}
  & T = T_\ast \left( \frac{r}{r_\ast} \right)^{-k_\mathrm{t}},  \\
  \label{eq-rho-power}
  & \rho = \rho_\ast \left( \frac{r}{r_\ast} \right)^{-k_\mathrm{d}},
\end{align}
where $r_\ast = r_\mathrm{in}$. Characteristic scales of temperature and density in these distributions depend on the parameters of the $\alpha$-model. For zone ``B'', according to \cite{Shakura1973A&A....24..337S}
\begin{align}
  \label{eq-temp-scale}
  & T_\ast = 8.3\cdot10^7 (\alpha m)^{-1/5} \dot{m}^{2/5} = 2.7\cdot10^7\,\text{К}  \;,
  & k_\mathrm{t} = 9/10,  \\
  \label{eq-rho-scale}
  & \rho_\ast = 0.76\,(\alpha m)^{-7/10} \dot{m}^{2/5} = 0.33\,\text{g$/$cm$^3$}  \;,
  & k_\mathrm{d} = 33/20.
\end{align}
The temperature in this expression is assumed to be  that inside the disk, but not at its surface. The factor \eqref{eq-midplane-surface-temperature} is taken into account already. For zone ``C''
\begin{align}
  & T_\ast = 2.9\cdot10^7 (\alpha m)^{-1/5} \dot{m}^{3/10} = 1.3\cdot 10^7\,\text{K}  \;,
  & k_\mathrm{t} = 3/4,  \\
  & \rho_\ast = 3.8\,(\alpha m)^{-7/10} \dot{m}^{11/20} = 1.0\,\text{g$/$cm$^3$},
  & k_\mathrm{d} = 15/8.
\end{align}
As it is seen, though the profiles in the zones ``B'' and ``C'' have different slopes, they are normalized to the similar temperature values. In order not to complicate the model, it will be implied below  that the disk consists of the ``B'' zone alone.

The range of allowed values of $T_\ast$ and $\rho_\ast$ is rather wide and is limited by the condition \eqref{eq-no-zone-a} only. If one assumes that $\alpha = 0.001$ is the lowest realistic value of the turbulence parameter, then, inserting $m = 55$ and $\dot{m} = \dot{m}_\mathrm{low}$ into \eqref{eq-temp-scale} and \eqref{eq-rho-scale}, one obtains  $T_\ast = 4.8\cdot 10^7$\,K, $\rho_\ast = 1.87$\,g$/$cm$^{3}$. These values of temperature and density should be regarded as the upper estimate of them. The lower estimate one obtains assuming $\alpha = 1$, $\dot{m} = 0.01$. In this case, $T_\ast = 5.9\cdot10^6$\,K, $\rho_\ast = 7.29\cdot 10^{-3}$\,g$/$cm$^{3}$.

Distribution of the angular momentum along the disk is close to Keplerian. At this, contribution of the gas pressure and radial component of the kinetic energy to the radial balance of forces is small (of the order of the square of the ratio of the disk thickness to radial scale \cite{Armitage2007astro.ph..1485A,Kurbatov2017ARep...61..475K}). However, one can not neglect the thermal structure of the disk, which determines intensity of the shock wave and is important for evaluation of the response of the disk to the loss of mass by accretor. On the contrary, radial velocity component seems to be not significant in this process.

Let summarize formulation of the problem. There is an axisymmetric accretion disk in the gravitational field of a point mass. Distribution of the thermodynamic quantities along the disk corresponds to the zone ``B'' of the Shakura and Sunyaev model. Centrifugal force in the disk completely balances gravity and gas pressure (radial velocity is equal to zero everywhere). At the initial moment, accretor mass decreases by $5\%$. In this state the disk is not in equilibrium. Further evolution of the disk state is calculated in the approximation of non-dissipative gas dynamics. We are interested in the solution for the midplane of the disk, while the vertical structure of the disk is neglected (justification for this will be presented in Conclusion). As initial conditions we accept distributions of temperature and density \eqref{eq-temp-power} --- \eqref {eq-rho-scale}.

In the next Section we will describe equations of the problem and present the procedure of their numerical solution.

\section{Numerical model}
\label{sec-num-model}

In the axisymmetric case, in the cylindrical coordinates, it is convenient to introduce the mass Lagrange coordinate $q$ instead of the radial coordinate $r$. The former satisfies relation $dq = r^2\rho\,dr$. As a result, the equations of gas dynamics describing the evolution of the accretion disk after the merger of black holes can be written as:
\begin{align}
  \label{eq-nm-1}
  & \diff{}{t}\left( \frac{1}{\rho} \right) = \pdiff{}{q} \left( r v_r \right),  \\
  \label{eq-nm-2}
  & \diff{v_r}{t} = -r\,\pdiff{P}{q} + \frac{v_\varphi^2}{r} - (1-\xi)\,\frac{GM}{r^2},  \\
  \label{eq-nm-3}
  & \diff{v_\varphi}{t} = -\frac{v_r v_\varphi}{r},  \\
  \label{eq-nm-4}
  & \diff{\varepsilon}{t} = -P\,\pdiff{}{q} \left( r v_r \right),  \\
  \label{eq-nm-5}
  & \diff{r}{t} = v_r.
\end{align}
Here, $\rho$ is density; $v_r$ --- radial velocity; $v_\varphi$ --- azimuthal velocity; $P$ --- pressure; $\varepsilon$ --- specific internal energy; $M$ --- total mass of binary black hole before the merger; $\xi$ --- the fraction of black holes mass emitted by gravitational waves. Let note that instead of the equation \eqref{eq-nm-1} one may use an equivalent equation
\begin{equation}
  \label{eq-nm-1a}
  \frac{1}{\rho} = r\,\pdiff{r}{q}.
\end{equation}
Additionally, from Eqs. \eqref{eq-nm-3} and \eqref{eq-nm-5} follows specific angular momentum $l = r v_\varphi$ conservation law
\begin{equation}
  \label{eq-nm-6}
  \diff{l}{t} = 0\,.
\end{equation}
In the region of the disk in which one may neglect radiation pressure compared to the gas pressure, equation of state is defined by relations
\begin{equation}
  \label{eq-nm-7a}
  P = P(\rho, T) = A \rho T,
\end{equation}
\begin{equation}
  \label{eq-nm-7b}
  \varepsilon
  = \varepsilon(\rho, T) = \frac{AT}{\gamma - 1},
\end{equation}
where $A = k_\mathrm{B}/m$; $k_\mathrm{B}$ --- Boltzmann constant; $m = 0.5 m_\mathrm{p}$ --- mean molecular mass; $m_\mathrm{p}$ --- proton mass; $\gamma = 5/3$ --- adiabatic exponent; mean molecular weight of the completely
ionized plasma is taken as $0.5$.

Let set the initial temperature and density distributions $T(r, 0)$ and $\rho(r, 0)$ to be power functions of radius, \eqref {eq-temp-power} and \eqref{eq-rho-power}. From these definitions and equations of state \eqref{eq-nm-7a}, \eqref{eq-nm-7b}, it is possible to obtain the corresponding distributions of pressure and specific internal energy:
\begin{equation}
  \label{eq-nm-10}
  P(r, 0)
  = \rho_\ast c_\ast^2 \left( \frac{r}{r_\ast} \right)^{-k_d-k_t},
\end{equation}
\begin{equation}
  \label{eq-nm-11}
  \varepsilon(r, 0)
  = \frac{c_\ast^2}{\gamma-1} \left( \frac{r}{r_\ast} \right)^{-k_t},
\end{equation}
where we use notation $c_\ast^2 = A T_\ast$. Assuming that initial radial velocity is equal to zero, from condition of the radial equilibrium in the disk
\begin{equation}
  \label{eq-nm-12}
  \frac{1}{\rho}\,\pdiff{P}{r}
  = - \frac{GM}{r^2} + \frac{l^2}{r^3}
\end{equation}
it is possible to obtain
\begin{equation}
  \label{eq-nm-13}
  l^2
  = r_\ast^2 c_\ast^2 \left[ \mu^2 \left( \frac{r}{r_\ast} \right)
      - (k_d + k_t) \left( \frac{r}{r_\ast} \right)^{2-k_t} \right]
\end{equation}
or
\begin{equation}
  \label{eq-nm-14}
  v_\varphi
  = c_\ast \left[ \mu^2 \left( \frac{r}{r_\ast} \right)^{-1}
      - (k_d + k_t) \left( \frac{r}{r_\ast} \right)^{-k_t} \right]^{1/2}  \;,
\end{equation}
where
\begin{equation}
  \label{eq-nm-15}
  \mu = \left( \frac{GM}{r_\ast c_\ast^2} \right)^{1/2}
  \equiv 165 \left( \frac{T_\ast}{10^7~\text{К}} \right)^{-1/2}  \;.
\end{equation}
In its sense, parameter $\mu$ is equal to the Mach number at the radius $r = r_\ast$.

For the further estimates we will also need characteristic semi-thickness of the disk. Let evaluate it as the vertical scale of the equilibrium disk:
\begin{equation}
  \label{eq-nm-17}
  H
  = \frac{r}{v_\phi}\,(AT)^{1/2}
  = r_\ast \left[ \mu^2 \left( \frac{r}{r_\ast} \right)^{k_t-3}
      - (k_d + k_t) \left( \frac{r}{r_\ast} \right)^{-2} \right]^{-1/2}  \;.
\end{equation}

For construction of the computational algorithm, it is convenient to use dimensionless variables. This is achieved by selection of the characteristic scales according to the scheme: $f \to f_ {0}f$, where $f$ is some sort value and $f_0$ is its dimensional scale. In order not to encumber the expressions, the dimensionless variables will be denoted by the same symbols as the original dimensional variables. As the scale dimensions, we choose the following values:
\begin{equation}
  \label{eq-nm-16}
  r_0 = r_\ast \;,\quad
  \rho_0 = \rho_\ast \;,\quad
  v_0 = c_\ast \;,\quad
  t_0 = r_\ast/c_\ast \;,\quad
  P_0 = \rho_\ast c_\ast^2 \;,\quad
  \varepsilon_0 = c_\ast^2 \;,\quad
  q_0 = r_\ast^2 \rho_\ast  \;,
\end{equation}

In the dimensionless variables, equation for radial velocity \eqref{eq-nm-2} and equations of state \eqref{eq-nm-7a}, \eqref{eq-nm-7b} will become:
\begin{equation}
  \label{eq-nm-2a}
  \diff{v_r}{t} = -r\,\pdiff{P}{q} + \frac{v_{\varphi}^2}{r} - \frac{(1-\xi)\,\mu}{r^2}  \;,
\end{equation}
\begin{equation}
  \label{eq-nm-7c}
  P = \rho T  \;,\qquad
  \varepsilon = \frac{T}{\gamma - 1}  \;.
\end{equation}
Remaining equations will not change their form after respective change of notation.

To solve these equations numerically, we used the implicit completely conservative Samarskii-Popov difference scheme \cite{Samarskii1980MoIzN....W....S}, some details of which are described in the Appendix. It should be emphasized that in this scheme not only the difference analogs of the laws of conservation of mass, momentum and energy are fulfilled, but additional relationships describing the balance for certain types of energy are fulfilled too. In addition, in our case the difference analogue of the law of conservation of the specific angular momentum \eqref{eq-nm-6} is fulfilled.

In the dimensionless formulation, the problem has only two parameters: $\xi$ and $\mu$. We set the fraction of the lost mass using the estimate obtained from gravitational-wave observations \cite{Abbott2017PhRvL.119n1101A}, $\xi = 0.05$. In the preceding Section, we have shown that for the realistic values of the turbulence parameter $\alpha$ and the dimensionless accretion rate $\dot{m}$, characteristic temperature  $T_\ast$ is $5.9\cdot 10^6 \mathdash 4.8\cdot 10^7$\,K. This interval corresponds to the range of Mach numbers $\mu = 75 \mathdash 215$. In numerical simulations, we set certain $\mu$ and from them, according to \eqref{eq-nm-15},
obtained $T_\ast$, $c_\ast$, and $t_\ast$.

\begin{figure}
  \centering
  \includegraphics{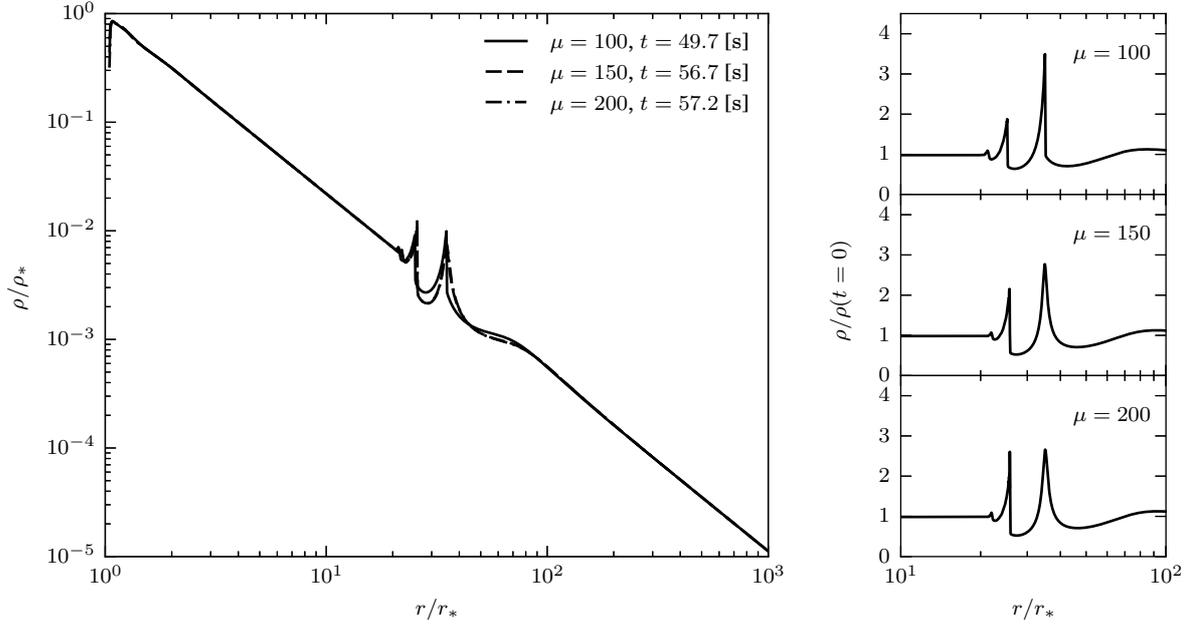}
  \caption{Radial density distribution in the models with different Mach number $\mu$. The moments of time are chosen in such a way that the positions of the shock wave coincide in all cases.}
  \label{fig-rho}
\end{figure}

\begin{figure}
  \centering
  \includegraphics{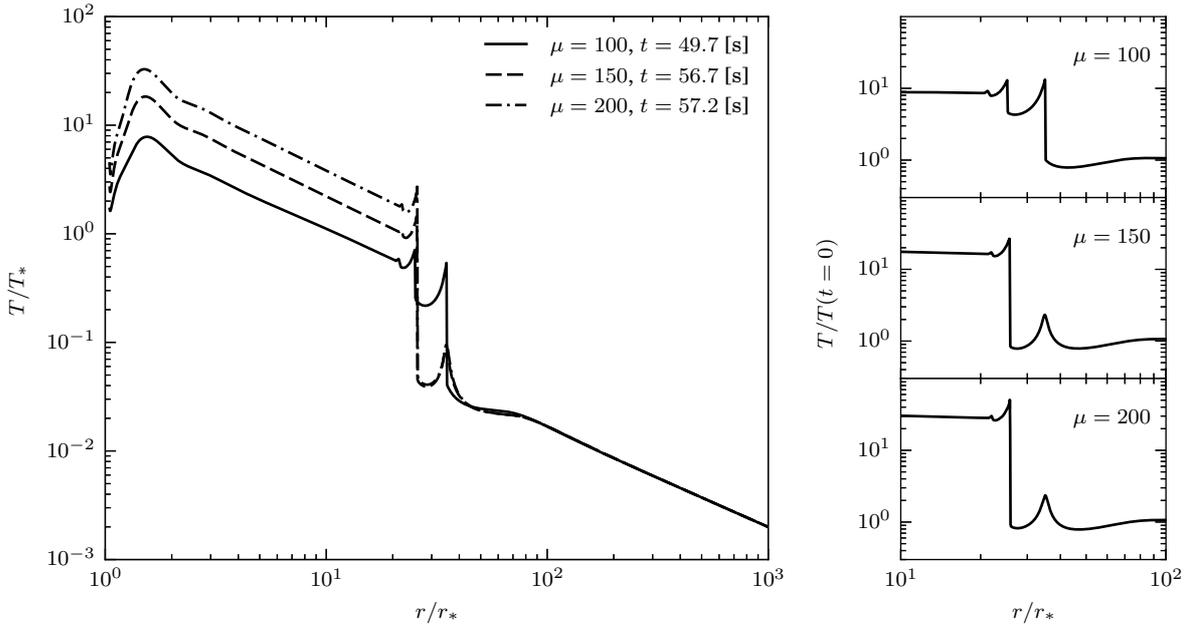}
  \caption{Radial distribution of the temperature.  Notation is the same as in Fig.~\ref{fig-rho}.}
  \label{fig-temp}
\end{figure}

\begin{figure}
  \centering
  \includegraphics{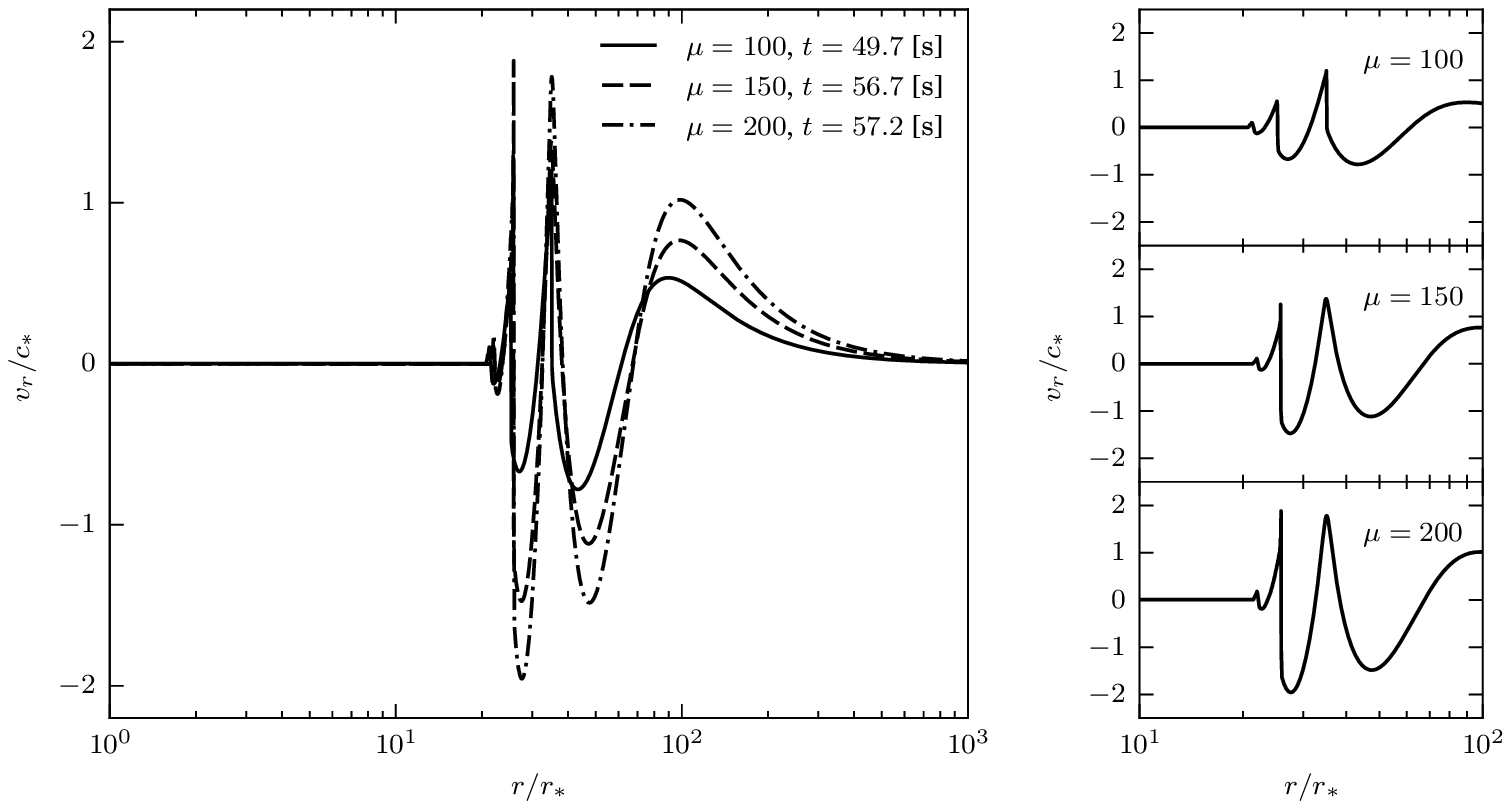}
  \caption{Radial distribution of the radial velocity. Notation is the same as in Fig.~\ref{fig-rho}.}
  \label{fig-v_r}
\end{figure}

\begin{figure}
  \centering
  \includegraphics{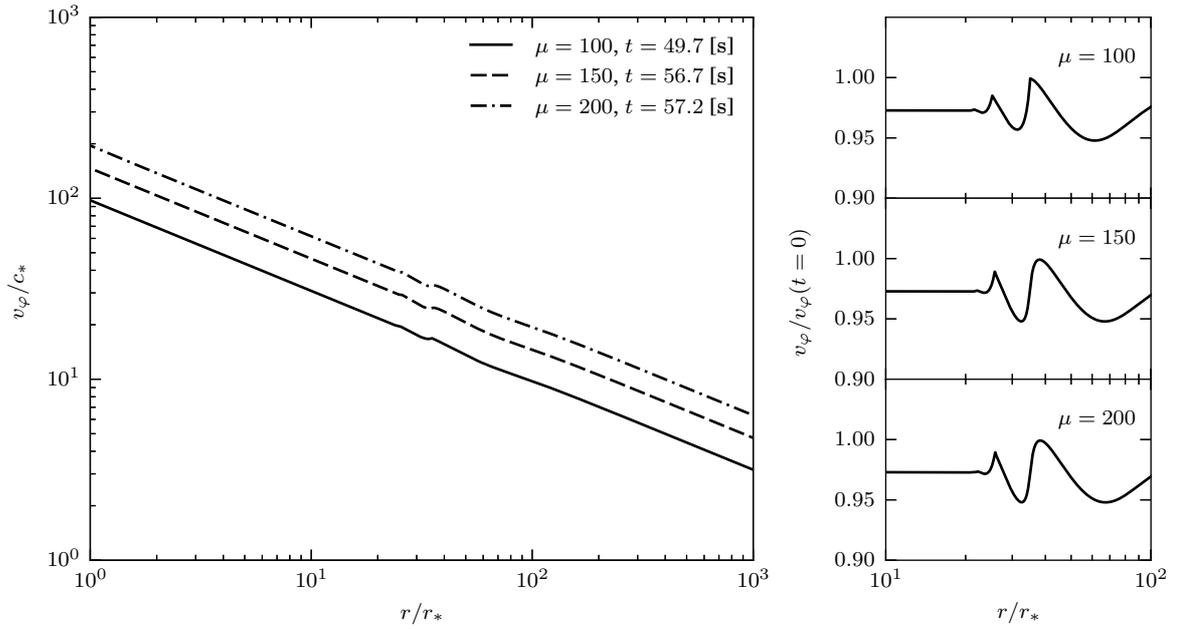}
  \caption{Radial distribution of the tangential velocity. Notation is the same as in Fig.~\ref{fig-rho}.}
  \label{fig-v_phi}
\end{figure}

We computed three numerical models: for $\mu = 100, 150, 200$. These values of $\mu$, as well as the fraction of the lost mass  $\xi$, correspond to the object GW170814 \cite{Abbott2017PhRvL.119n1101A} --- the latest, at the time of writing, detected event of merger of a binary BH. The plots of dimensionless density, temperature, radial and tangential velocities are shown in Figs.~\ref{fig-rho}--\ref{fig-v_phi}. All calculations revealed qualitatively similar properties of the flow: one or two shock waves appear in the central region of the disk, depending on $\mu$. The shock propagates through the disc slowing down. The density at the front of the shock wave jumps, changing three or more times, the temperature changes, approximately, by a factor
from $7.5$ to $27$, compared to the initial value (see Table~\ref{tbl-shock}). Radial velocity oscillates with an amplitude of the order of the speed of sound and a larger one. Tangential velocity varies little. Characteristics of the flow in the front of the shock remain practically constant over computations time.

\begin{table}
  \centering
  \begin{tabular}{rrr}
    \hline
    $\mu$ & $T_\ast$~[К] & $T/T(t=0)$  \\
    \hline
    $100$ & $2.7\cdot10^7$ & $7.5$  \\
    $150$ & $1.2\cdot10^7$ & $15.5$  \\
    $200$ & $0.43\cdot10^7$ & $27$
  \end{tabular}
  \caption{Some characteristics of the disks. In the columns are listed Mach number, the temperature in the central region of unperturbed
disk, temperature discontinuity at the shock.}
  \label{tbl-shock}
\end{table}

The flow of gas in the perturbed region of the disk is a combination of density, temperature, and radial velocity waves of large amplitude. For $\mu = 200$ and $150$, one shock wave with a precursor (a caustic in the density and temperature distribution) appears at the outer boundary of the perturbed region.  For $\mu = 100$ the caustic turns into a shock wave. We also carried out calculations for Mach numbers $\mu = 50$ and $\mu = 10 $ (they are not shown in the Figures). It is interesting that for small values of $\mu$ there is only one shock wave with a precursor in the disk.

\begin{figure}
  \centering
  \includegraphics{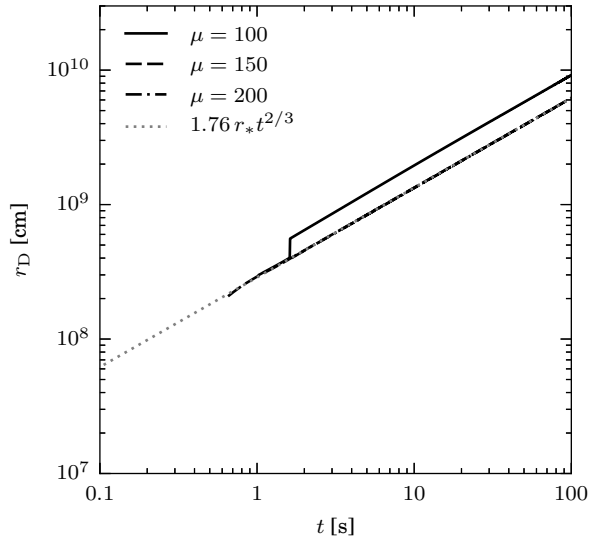}
  \caption{Position of the shock wave as a function of time. Approximate relation is shown by gray dotted line.}
  \label{fig-shock}
\end{figure}

It turned out that the position of the shock wave as a function of time does not depend on the Mach number in the disk, $\mu$ and admits a simple approximation
\footnote{%
The discrepancy between the plots for $\mu = 100$ and other ones is due to the fact that in the program code the shock is searched for from the right (external) part of the computational domain. At the same time, for this Mach number there are two shock waves in the flow (see Fig.~\ref{fig-temp}) and the position of the internal shock coincides with the position of the shock for $\mu = 150$ and $\mu = 200$.}
(Fig.~\ref{fig-shock}):
\begin{equation}
  \label{eq-shock-position}
  r_D = 1.76\,r_\ast t^{2/3}  \;.
\end{equation}
From this expression one may obtain velocity of the shock:
\begin{equation}
  \label{eq-shock-speed}
  D = \diff{r_D}{t}
  = 1.2\,r_\ast t^{-1/3}
  = 1.59\,r_\ast \left( \frac{r_D}{r_\ast} \right)^{-1/2}  \;.
\end{equation}

\section{The spectrum of electromagnetic radiation and light-curves}
\label{sec-em-spectra}

Shock wave arising in the disk leads to the temperature increase by $7.5\mathdash 27$ times, depending on the Mach number (Fig.~\ref{fig-temp}, Table~\ref{tbl-shock}). As a result, the gas in the inner region of the disk warms up to the temperature of approximately $2\cdot10^8$\,K, which corresponds to the thermal energy per particle of about $17.2$~keV. Under conditions of a nonmagnetic accretion disk, the spectrum of electromagnetic radiation in this energy band is determined by two processes \cite{Ginzburg1979tpa..book.....G}: bremsstrahlung of electrons and Compton scattering. The latter can lead to an increase of photon energy and distortion of the spectrum --- a decrease of the intensity at the Planck maximum frequencies and appearance of a ``tail'' in the high-energy region \cite{Illarionov1972SvA....16...45I,Pavlov1989PhR...182..187P}. However, comptonization of the spectrum is effective only if the radiation travels through a transparent layer \cite {Ginzburg1979tpa..book.....G,Pavlov1989PhR...182..187P}. Let us estimate the optical thickness of the disk:
\begin{equation}
  \tau_\nu = (\sigma_\mathrm{T} + \sigma_\mathrm{ff,\nu})\,\Sigma  \;.
\end{equation}

Substituting  expression for the absorption cross-sections, for instance, from \cite{Illarionov1972SvA....16...45I}, it can be shown that in the central region of the disk ($\rho \sim 1$\,g$/$cm$^3$, $T \sim 10^8$\,K, $H \sim 10^6$\,cm), optical thickness in the entire spectrum is much larger than $1$ and at the wavelengths shorter than $10$\,nm it is determined by Compton scattering. This  is characteristic for the outer regions of the disk too. Distortion of the spectrum in the optically thin layer close to the disk surface is also unimportant, since the ratio of the geometric thickness of this layer and the thickness of the disk (it can be estimated as $\tau_\nu^{-1}$) is negligible at all frequencies. Thus, it can be assumed that the radiation from the surface of the disk has a Planck spectrum%
\footnote{%
Comptonization of electromagnetic radiation is effective in the zone ``A'' of the standard $\alpha$-disk
\cite{Shakura1973A&A....24..337S}. In the disk model we use, zone ``A'' is absent, while in the zone
``B'' surface density is by $4\mathdash 5$ orders of magnitude larger than the estimate for the zone ``A''.}:
\begin{equation}
  B_\nu
  = \frac{2h\nu^3}{c^2}\,\frac{1}{e^{h\nu/k_\mathrm{B} T_\mathrm{s}} - 1}  \;,
\end{equation}
where $T_\mathrm{s}$ is surface temperature (see \S~\ref{sec-setting} and relation \eqref{eq-midplane-surface-temperature}).

\begin{figure}
  \centering
  \includegraphics{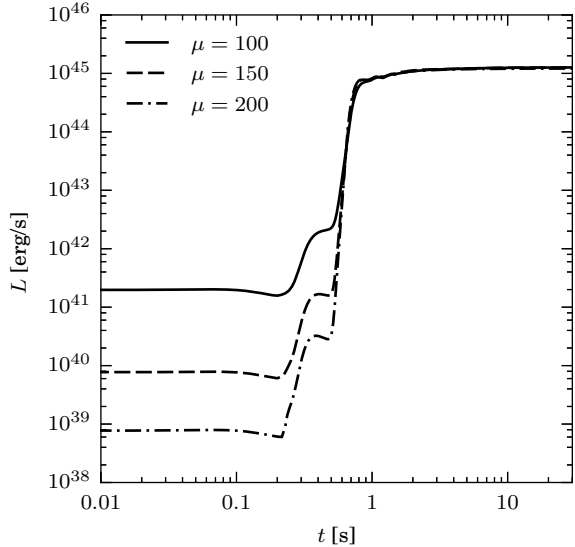}
  \caption{Bolometric light curves in the  models with different Mach numbers.}
  \label{fig-light_curve}
\end{figure}

In Fig. \ref{fig-light_curve} we show variation of disk bolometric luminosity with time, computed as
\begin{equation}
  L(t) = 2\pi \int_{r_\ast}^\infty dr\,r\,\sigma_\mathrm{SB} T_\mathrm{s}^4(t, r)  \;.
\end{equation}
It is remarkable that different values of the parameter $\mu$ led, ultimately, to the same luminosity of the disk, about $10^{45}$\,erg$/$s. Analytically, luminosity can be estimated as $L \sim \pi r_\ast^2 \sigma_\mathrm{SB} f^4 T_\ast^4$, where $f$ is the jump of temperature at the shock (see the third column of Table~\ref{tbl-shock}). Using the estimate of the maximum luminosity and applying Eq.~\eqref{eq-nm-15}, we obtain
\begin{equation}
  f = 27 \left( \frac{\mu}{200} \right)^2  \;.
\end{equation}

Let recall that in this study we are based on the hydrodynamic model, without account of radiative processes. In Figs.~\ref{fig-temp} and \ref{fig-light_curve} it is seen that, after the jump at the shock gas temperature practically does not change. This approximation is certainly not fair for long timescales, because  radiative cooling should lead to the cooling of the disk after the passage of the shock wave and gradual setting to a new equilibrium state, in which the rate of de-excitation will be equal to the rate of viscous heating. Let estimate the time of radiative cooling as the time of radiation diffusion:
\begin{equation}
  t_\mathrm{rad} \sim (\sigma_\mathrm{T} + \sigma_\mathrm{ff})\,\Sigma\,\frac{H}{c}  \;,
\end{equation}
where $\sigma_\mathrm{T} = 0.40$\,cm$^2/$g and $\sigma_\mathrm{ff} = 0.11 N/T^{7/2}$\,cm$^2/$g; $N$ --- electron number density; $\Sigma = 2\rho H$. For instance, in the vicinity of the inner disk radius $\rho \sim 1$\,g$/$cm$^3$, $T \sim 10^8$\,K, and $H \sim 10^6$\,cm. This gives $t_\mathrm{rad} \approx 31.6$\,s. A more general expression has the form (at such temperatures cross-section $\sigma_\mathrm{ff}$ does not contribute to the absorption)
\begin{equation}
  \label{eq-t_rad}
  t_\mathrm{rad} = \frac{2 \sigma_\mathrm{T} \rho_\ast H_\ast^2}{c} \left( \frac{r}{r_\ast} \right)^{3 - (k_d+k_t)}  \;.
\end{equation}
Characteristic width of the hot region behind the shock may be estimated as
\begin{equation}
  \label{eq-hot-width}
  \Delta r_D = D t_\mathrm{rad}  \;,
\end{equation}
where $D$ is velocity of the shock wave \eqref{eq-shock-speed}. After insertion of $r_D$ instead of $r$ into expressions \eqref{eq-t_rad} and \eqref{eq-hot-width}, it is seen, that the width of the hot region $\Delta r_D$ weakly depends on time (as $t^{-0.075}$) and is
\begin{equation}
  \Delta r_D \approx 42.4\,r_\ast  \;.
\end{equation}
This means that the radiative cooling should not strongly change temperature distribution presented in Fig.~\ref{fig-temp}. The time of decrease of the temperature to its initial value we estimate as:
\begin{equation}
  t_\mathrm{cool} = t_\mathrm{rad}\,\frac{T}{T(t=0)}  \;.
\end{equation}
If the temperature behind the shock increases by factor $7.5 \mathdash 27$, the decline time of the light-curve will be $4$ to $14$ min.

\begin{figure}
  \centering
  \includegraphics{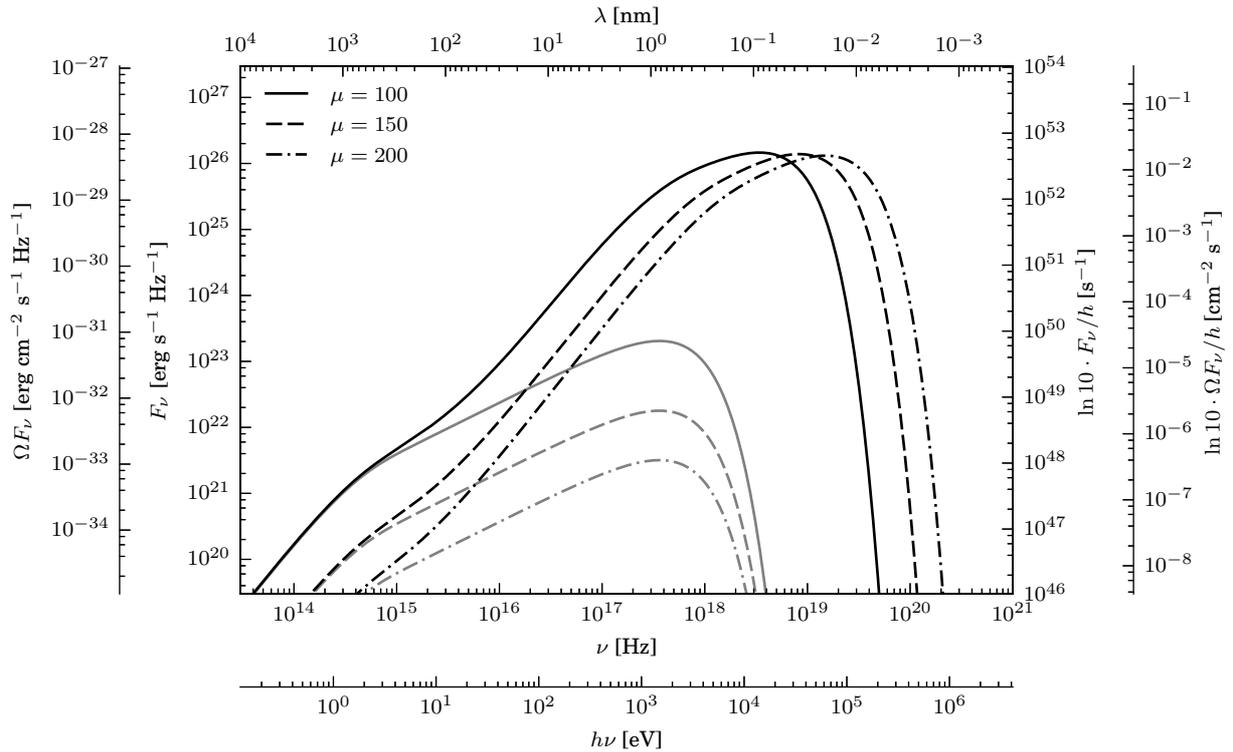}
  \caption{The spectrum of electromagnetic radiation. Inner ordinates show the scales of spectral flux density (left axis) and the estimate of the photon flux per unit logarithmic frequency interval (right axis). Outer y-axes correspond to the source distance of $540$~Mpc. They show the scales of spectral density of the flux through a unit area (axis to the left) and the scale of the estimate of the photon flux through a unit area (axis to the right). In gray are shown the spectra of unperturbed disks for the corresponding values of $\mu$.}
  \label{fig-spectra}
\end{figure}

Spectral density of the emission flux from one side of the disk in the perpendicular direction is
\begin{equation}
  F_\nu
  = 2\pi^2 \int_{r_\ast}^\infty dr\,r B_\nu[T_\mathrm{s}(r)]  \;.
\end{equation}
It is seen in Fig.~\ref{fig-spectra} that the main fraction of the energy is emitted in X-ray and gamma-ranges.

For gamma-range it is convenient to use the estimate of the number of photons, emitted per unit of time in the unit logarithmic frequency range (right scale in Fig.~\ref{fig-spectra}):
\begin{equation}
  \label{eq-photon-flux}
  \int_\nu^{10\nu} d\nu'\,\frac{F_{\nu'}}{h\nu'}
  = \ln 10 \int_\nu^{10\nu} d(\lg \nu')\,\frac{F_{\nu'}}{h}
  \sim \ln 10\,\frac{F_\nu}{h}  \;.
\end{equation}
The outer y-axes show the flux through a $1$\,cm$^2$ area under condition that the observed object is at the distance of $540$\,Mpc (the system GW170814 \cite{Abbott2017PhRvL.119n1101A}). For this, the flux was corrected by the factor
\begin{equation}
  \Omega = \frac{1}{(540\,\text{Mpc})^2} = 3.6\cdot10^{-55}\,\text{cm$^{-2}$}  \;.
\end{equation}

\begin{figure}
  \centering
  \includegraphics{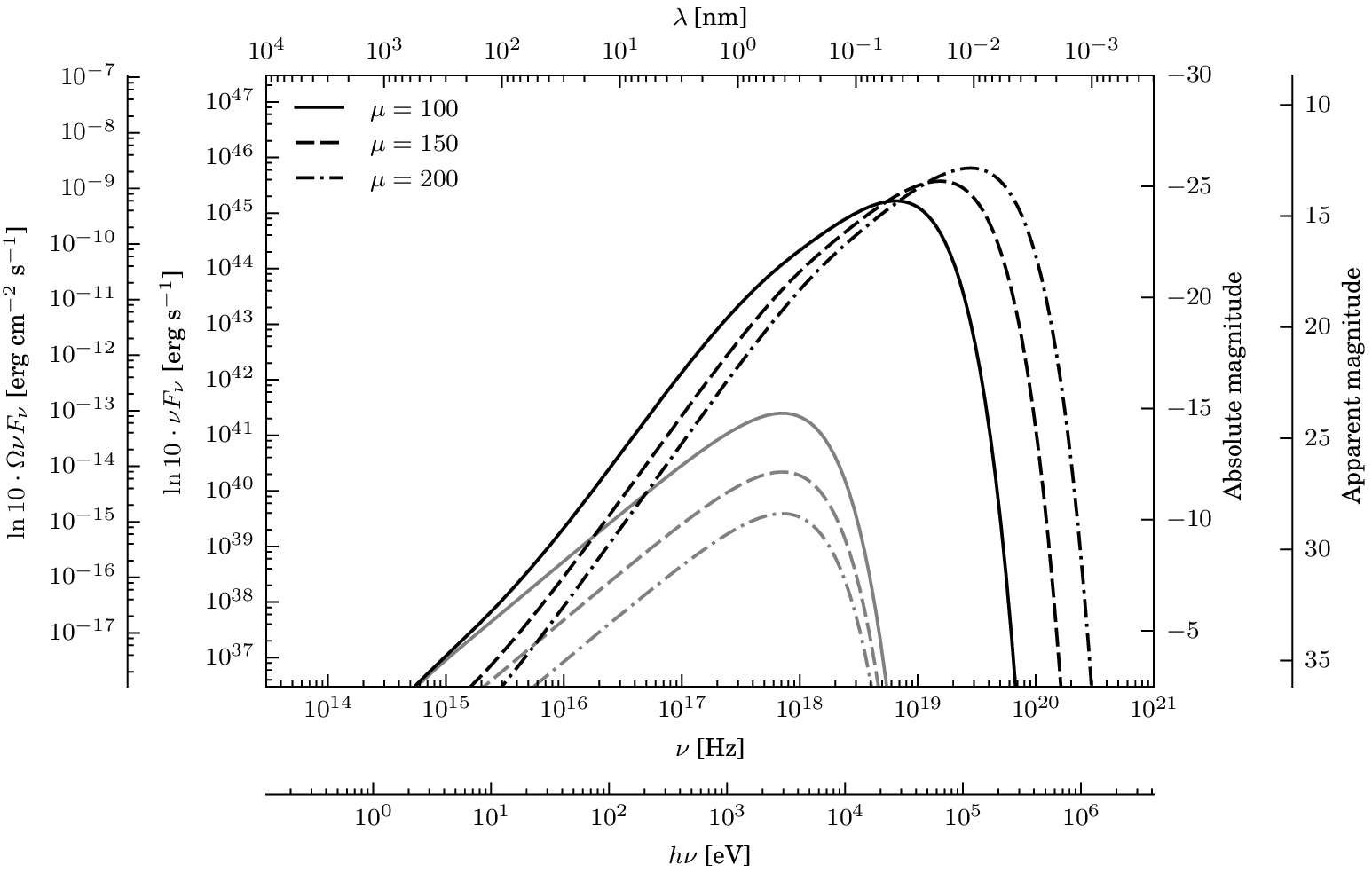}
  \caption{Estimate of the luminosity per unit logarithmic frequency interval. Inner plot ordinates show luminosity (left axis) and absolute bolometric stellar magnitude (right axis). Outer y-axes correspond to the source distance of $540$~Mpc. Axis to the left shows the scale of the flux through a unit detector area, axis to the right shows the scale of the apparent stellar magnitude. Gray lines show the spectra of unperturbed disks for the corresponding values of $\mu$.}
  \label{fig-magnitudes}
\end{figure}

In a similar way one may compute approximately the energy, emitted per unit of time in the unit logarithmic frequency range (Fig.~\ref{fig-magnitudes}):
\begin{equation}
  \int_\nu^{10\nu} d\nu'\,F_{\nu'}
  \sim \ln 10\cdot\nu F_\nu  \;.
\end{equation}
This value may be considered as the estimate of luminosity in the one order of magnitude wide frequency band.

Figures \ref{fig-spectra} and \ref{fig-magnitudes} show that the main fraction of energy is emitted in the intermediate and hard X-ray ($1 \mathdash 100$\,keV) and gamma-rays ($\gtrsim 100$\,keV).

In the intermediate X-ray range, up to $10$\,keV, luminosity of the gas heated by the shock wave is $10^{43} \mathdash 10^{45}$\,erg$/$s, exceeding by $2$ to $4$ orders of magnitude (depending on $\mu$) luminosity of the unperturbed disk, which, in turn, is comparable to the luminosity of X-ray pulsars \cite{Lipunov1992ans..book.....L} and ultraluminous X-ray sources \cite{Feng2006ApJ...650L..75F,Fabrika2017ASPC..510..395F}. After correction for the distance to GW170814 ($540$\,Mpc \cite{Abbott2017PhRvL.119n1101A}), observed flux from the perturbed disk in this spectral range appears to be  $\gtrsim 10^{-9}$\,erg$/$cm$^2/$s, several orders of magnitude larger than the sensitivity threshold of the EPIC instrument of the XMM-Newton observatory ($5\sigma$ level for exposition $10^5$\,s) \cite {XMM-Newton2018UsersHandbook}.

Radiation flux of $10^{45}$\,erg$/$s and higher, at the frequencies $10 \mathdash 300$\,keV, can provide ``hot'' $\alpha$-disks ($\sim 10^9$\,K) around stellar-mass BH \cite{Pavlov1989PhR...182..187P,Maraschi1990ApJ...353..452M} or an accreting supermassive BH. Observed flux density in the hard X-ray range will be only one or two orders of magnitude lower than that of the Crab nebula ($1 \text{\,Crab} = 2.4\cdot10^{-8}$\,erg$/$cm$^2/$s).

In the gamma-range, radiation flux is many orders of magnitude larger than the flux from the unperturbed disk and it is comparable to the flux from soft gamma-sources, $\gtrsim 10^{26}$\,erg$/$s$/$Hz \cite{Starling2011MNRAS.411.2792S}. At a wavelength of $100$\,keV,  observed flux is $10^{-2}$\,photons$/$cm$^2/$s per unit of logarithmic wavelength interval (dex), or $10^{-4 }$\,photons$/$cm$^2/$s$/$keV. Let note that the limiting sensitivity of the INTEGRAL instrument IBIS at this wavelength is $2.85\cdot10^{-6}$\,photons$/$cm$^2/$s$/$keV ($3\sigma$-level for the exposition $10^5$\,s) \cite {Ubertini2018IBIS,Ubertini2003A&A...411L.131U}.

In the EUV, $10 \mathdash 10^2$\,eV, the brightening of the disk may exceed an order of magnitude or $2.5^\mathrm{m}$ and reach $3\cdot 10^{40} \mathdash 3\cdot10^{41}$\,erg$/$s. In FUV, the brightening does not exceed factor $2$. At larger wavelength the brightening is practically absent. In the near-UV range the brightness of the disk is by $4 \mathdash 5$ orders of magnitude lower than the brightness of a typical supernova ($10^{41}$\,erg$/$s).

Luminosity estimates presented above are very approximate also because of uncertain distance to the binary system. The minimum and maximum estimates of the distance to  GW170814 \cite{Abbott2017PhRvL.119n1101A} differ by factor 2 (Table~\ref{tbl-sources}). To the variation of distance by a factor two corresponds the change of the apparent brightness by a factor of four or the change of the visual magnitude by $1.5^\mathrm{m}$.

\section{Discussion and conclusion}
\label{sec-conclusions}

In the numerical model described in \S~\ref{sec-num-model}, we neglected  vertical structure of the disk. Now, using results of computations, we can check the validity of this approximation. After a fraction $\xi$ of BH mass is lost, the matter of the disk acquires an acceleration in the vertical direction $\dot{v}_z \sim \xi H \Omega^2$. Characteristic disk response time is
\begin{equation}
  t_z
  \sim \frac{f^{1/2} c_\mathrm{s}}{\dot{v}_z}
  \sim t_\ast\,\frac{f^{1/2}}{\xi\mu} \left( \frac{r}{r_\ast} \right)^{3/2}  \;,
\end{equation}
where $f$ is the temperature increase factor (see the previous section). We used here the definitions from \S~\ref{sec-num-model}. The time $t_z$ should be compared to the propagation time of the shock wave, $t_D \sim r/D$. Taking into account the estimates of $f$ and $D$ from \S~\ref{sec-em-spectra}, we obtain
\begin{equation}
  \frac{t_z}{t_D}
  \sim \frac{40}{\xi\mu}  \;.
\end{equation}
As one can see, for $\xi = 0.05$ and $\mu \sim$ 100 disk relaxation in the vertical direction does not affect the propagation of the shock wave. It can be shown, however, that the time $t_z$ turns out to be of the same order of magnitude as the time $t_\mathrm{cool}$. Therefore, computation of radiation at larger times will require a more complete account of the vertical structure of the disk.

In the present study, we computed the response of the circumbinary accretion disk to the decrease of the mass of a binary BH. The idea was that the merging black holes lose about $5\,\%$ of the total mass-energy via radiation of gravitational waves. As a result, the matter of the initially equilibrium quasi-Keplerian disk acquires an excess of momentum in the direction from the accretor. The impulse is caused by the disbalance of centrifugal force and reduced gravitational force of  accretor. This difference is larger in the inner part of the disk, leading to the propagation of a strong perturbation over the disk.

We applied these considerations to the object GW170814 --- a binary BH with a total mass of about $55\,M_\odot$, located at the distance of $540$\,Mpc. As a model of the circumbinary disk, the standard model of Shakura and Sunyaev was taken. Gas pressure prevailed in the disk, for this reason the initial temperature of the disk was assumed to be rather low, $\lesssim 3 \cdot10^7$\,K. We found that out of the inner region of the accretion disk  begins to spread a perturbation, which rapidly becomes a shock wave. The temperature behind the shock jumps by a factor $7.5$ to  $27$, depending on the initial temperature of the disk.

Computation of the bolometric light curves showed that the luminosity of the disk increases by $4 \mathdash 6$ orders of magnitude, up to $10^{45}$\,erg$/$s, which corresponds to the absolute stellar magnitude $-23.8^\mathrm{m}$. The maximum of the radiation flux is in the X-ray and gamma ranges. In the spectral band of the EPIC instrument of the XMM-Newton observatory or the eROSITA telescope
of the Spectrum-RG observatory, luminosity increase, compared to the unperturbed disk, is by $3 \mathdash 4$ orders of magnitude ($7.5^\mathrm{m} \mathdash 10^\mathrm{m}$ in absolute magnitudes); this corresponds to the apparent magnitude of $12^\mathrm{m}$. In the observational band of the IBIS telescope of the INTEGRAL observatory, luminosity is at maximum and corresponds to the apparent flux of $10^{-4}$\,photons$/$cm$^2/$s$/$keV at the wavelength $\sim100$\,keV. In the FUV-range, and toward higher wavelengths, the perturbation of the disk practically does not lead to its brightening: at a wavelength of $10$\,eV, the luminosity increases approximately twice and corresponds to the apparent stellar magnitude of $32^\mathrm{m}$. The estimates of visible fluxes in the X-ray range exceed the sensitivity threshold of the EPIC instrument by several orders of magnitude (up to four, depending on the exposition). The flux in gamma range is about one and a half orders higher than the sensitivity threshold of IBIS aboard INTEGRAL observatory.

For realistic parameters of the accretion disk, estimated time of the light curve decline is of the order of several minutes. It is interesting that the bolometric luminosity does not depend on the initial temperature of the disk, but is determined, apparently, by the mass of accretor and the fraction of the mass lost. At this, the shape of the spectrum is sensitive to the temperature: an increase in the initial temperature of the disk leads to the shift of the spectrum into the short-wavelength region.

In the forthcoming papers we plan to investigate further the effect of mass loss by merging black holes applying a more sophisticated model of accretion disks with account of radiation pressure and gas cooling.

The authors acknowledge A.V. Tutukov, Ya.N. Pavlyuchenkov and O.Yu Malkov for helpful comments.

\section*{Appendix. Difference scheme}

Let define in the computational domain a difference mesh $q_i$, $i = 0, 1, \ldots, N$, with coordinates of the nodes related to the  Lagrange mass coordinate $q$. The step of the mesh $\Delta q_{i+1/2} = q_ {i+1} - q_ {i} $ has been chosen in such a way that in the logarithmic scale the distances between neighboring nodes is the same, i.e., $\ln (q_{i+1}/q_{i}) = \text{const}$. Radial coordinate and velocity at time instants $t^n$ will be referred to the mesh nodes: $r_i^n$, $v_{r, i}^n$, $v_{\varphi,i}^n$. The thermodynamic quantities will refer to the centers of the cells that are numbered by half-integral indices: $\rho^n_{i+1/2}$, $P^n_{i+1/2}$, $\varepsilon^n_{i+1/2}$, $T^n_{i+1/2}$ \cite{Samarskii1980MoIzN....W....S}. To simplify further records, let use the following notation for the operators of finite differences:
\begin{equation}\label{eq-a-1}
 (\Delta_t f)^{n+1/2} = \frac{f^{n+1} - f^n}{\Delta t}  \;,\qquad
 (\Delta_q f)_{i+1/2} = \frac{f_{i+1} - f_i}{\Delta q_{i+1/2}}  \;,\qquad
 (\Delta_q f)_{i} = \frac{f_{i+1/2} - f_{i-1/2}}{\Delta q_{i}}  \;,
\end{equation}
where $\Delta q_{i} = (\Delta q_{i+1/2} + \Delta q_{i-1/2})/2$. At this, if this will not arise misunderstanding, we will skip the indices of operators.

Equations \eqref{eq-nm-1a}, \eqref{eq-nm-2a}, \eqref{eq-nm-5}, \eqref{eq-nm-4}, \eqref{eq-nm-7c} may be approximated by the following difference scheme:
\begin{align}
  \label{eq-a-2}
  & \frac{1}{\rho^{n+1}_{i+1/2}}
  = \frac{1}{2}\,\Delta_q (r^{n+1})^2  \;,  \\
  \label{eq-a-3}
  & \Delta_t v_{r,i}
  = - r^{n+1/2}_i \Delta_q \Pi^{n+\sigma}
    + \frac{(v^{n+1/2}_{\varphi,i})^2}{r^{n+1/2}_i}
    - \frac{(1-\xi) \mu^2}{r^{n+1}_i r^n_i}  \;,  \\
  \label{eq-a-4}
  & \Delta_t r_i
  = v^{n+1/2}_{r,i}  \;,  \\
  \label{eq-a-5}
  & \Delta_t \varepsilon_{i+1/2}
  = - \Pi^{n+\sigma}_{i+1/2}\,\Delta_q \left( r^{n+1/2} v_r^{n+1/2}\right)  \;,  \\
  \label{eq-a-6}
  & \Delta_t v_{\varphi,i}
  = -\frac{v^{n+1/2}_{r,i} v^{n+1/2}_{\varphi,i}}{r^{n+1/2}_i}  \;,  \\
  \label{eq-a-7}
  & P^{n+1}_{i+1/2}
  = \rho^{n+1}_{i+1/2}\,T^{n+1}_{i+1/2}  \;,  \\
  \label{eq-a-8}
  & \varepsilon^{n+1}_{i+1/2}
  = \frac{T^{n+1}_{i+1/2}}{\gamma-1}  \;,  \\
  \label{eq-a-9}
  & \Pi^{n+1}_{i+1/2}
  = P^{n+1}_{i+1/2} + \omega^{n+1}_{i+1/2}  \;,  \\
  \label{eq-a-10}
  & \omega^{n+1}_{i+1/2}
  = \Omega \left( \rho^{n+1}_{i+1/2} \,,\: r^{n+1}_i \,,\: r^{n+1}_{i+1} \,,\:
      v^{n+1}_i \,,\: v^{n+1}_{i+1} \right)  \;,  \\
\end{align}
with notation
\begin{equation}
  \label{eq-a-11}
  \Pi^{n+\sigma}_{i+1/2}
  = \sigma \Pi^{n+1}_{i+1/2} + (1-\sigma)\,\Pi^{n}_{i+1/2}  \;.
\end{equation}
The parameter $\sigma$ characterizes the degree of implicitness of the scheme and varies from $0$ to $1$. This parameter determines the order of approximation over time. In the case of $\sigma = 1/2$ the scheme has the second order of approximation over $\Delta t$ and in the remaining cases it has the first order. The value $\omega$ describes artificial viscosity which is necessary for a more accurate description of solutions with shock waves and contact discontinuities. Explicit form of the function $\Omega$ is determined by the specific model of artificial viscosity. In our calculations, we used linear viscosity of the form
\begin{equation}\label{eq-a-12}
 \omega = -\frac{\eta \rho}{2}
 \left( \pdiff{v_r}{r} - \left| \pdiff{v_r}{r} \right| \right),
\end{equation}
where coefficient $\eta_{i+1/2} = \eta_0 \Delta q_{i+1/2}$. The value of $\eta_0$ was set to three to five.

This scheme is completely conservative in the sense that it provides not only the balance of the total energy, but of its individual types too (thermal, kinetic, rotational, and gravitational). In addition, it is not difficult to show that in this scheme keeps the difference relation
\begin{equation}\label{eq-a-13}
 \Delta_t \left( r_i v_{\varphi,i} \right) = 0,
\end{equation}
which describes angular momentum conservation law. This means that there exists an equality
\begin{equation}\label{eq-a-14}
 r^n_i v^n_{\varphi,i} = r^0_i v^0_{\varphi,i} = l_i,
\end{equation}
where $l_i$ is the  mesh function for the specific angular momentum. Then it follows that
\begin{equation}\label{eq-a-15}
 v^n_{\varphi,i} = \frac{l_i}{r^n_i}.
\end{equation}

The system of algebraic equations \eqref{eq-a-2}--\eqref{eq-a-10} is non-linear. For its solution a combination of Newton and elimination methods was applied.


\clearpage

\begin{thebibliography}{99}

\bibitem{Abbott2016ApJ...818L..22A}
B.~P. {Abbott}, R.~{Abbott}, T.~D. {Abbott}, M.~R. {Abernathy}, F.~{Acernese},
  K.~{Ackley}, C.~{Adams}, T.~{Adams}, P.~{Addesso}, R.~X. {Adhikari}, and
  et~al,
\newblock {Astrophys. J. Lett.}, {\bf 818}, L22 (2016).

\bibitem{Abbott2016PhRvL.116x1103A}
B.~P. {Abbott}, R.~{Abbott}, T.~D. {Abbott}, M.~R. {Abernathy}, F.~{Acernese},
  K.~{Ackley}, C.~{Adams}, T.~{Adams}, P.~{Addesso}, R.~X. {Adhikari}, and
  et~al,
\newblock {Physical Review Letters}, {\bf 116(24)}, 241103 (2016).

\bibitem{Abbott2017PhRvL.118v1101A}
B.~P. {Abbott}, R.~{Abbott}, T.~D. {Abbott}, F.~{Acernese}, K.~{Ackley},
  C.~{Adams}, T.~{Adams}, P.~{Addesso}, R.~X. {Adhikari}, V.~B. {Adya}, and
  et~al,
\newblock {Physical Review Letters}, {\bf 118(22)}, 221101 (2017).

\bibitem{Abbott2017ApJ...851L..35A}
B.~P. {Abbott}, R.~{Abbott}, T.~D. {Abbott}, F.~{Acernese}, K.~{Ackley},
  C.~{Adams}, T.~{Adams}, P.~{Addesso}, R.~X. {Adhikari}, V.~B. {Adya}, and
  et~al,
\newblock {Astrophys. J. Lett.}, {\bf 851}, L35 (2017).

\bibitem{Abbott2017PhRvL.119n1101A}
B.~P. {Abbott}, R.~{Abbott}, T.~D. {Abbott}, F.~{Acernese}, K.~{Ackley},
  C.~{Adams}, T.~{Adams}, P.~{Addesso}, R.~X. {Adhikari}, V.~B. {Adya}, and
  et~al,
\newblock {Physical Review Letters}, {\bf 119(14)}, 141101 (2017).

\bibitem{Tutukov2017ARep...61..833T}
A.~V. {Tutukov} and A.~M. {Cherepashchuk},
\newblock {Astronomy Reports}, {\bf 61}, 833 (2017).

\bibitem{Kaigorodov2010ARep...54.1078K}
P.~V. {Kaigorodov}, D.~V. {Bisikalo}, A.~M. {Fateeva}, and A.~Y. {Sytov},
\newblock {Astronomy Reports}, {\bf 54}, 1078 (2010).

\bibitem{Sytov2011ARep...55..793S}
A.~Y. {Sytov}, P.~V. {Kaigorodov}, A.~M. {Fateeva}, and D.~V. {Bisikalo},
\newblock {Astronomy Reports}, {\bf 55}, 793 (2011).

\bibitem{Shakura1973A&A....24..337S}
N.~I. {Shakura} and R.~A. {Sunyaev},
\newblock {Astron. Astrophys.}, {\bf 24}, 337 (1973).

\bibitem{Bode2007APS..APR.S1010B}
N. {Bode} and S. {Phinney},
\newblock {APS April Meeting Abstracts}, S1.010 (2007).

\bibitem{Bekenstein1973ApJ...183..657B}
J.~D. {Bekenstein},
\newblock {Astrophys. J.}, {\bf 183}, 657 (1973).

\bibitem{Kocsis2008PhRvL.101d1101K}
B. {Kocsis} and A. {Loeb},
\newblock {Physical Review Letters}, {\bf 101(4)}, 041101 (2008).

\bibitem{Megevand2009PhRvD..80b4012M}
M. {Megevand}, M. {Anderson}, J. {Frank}, E.~W. {Hirschmann}, L. {Lehner}, S.~L. {Liebling}, P.~M. {Motl}, and D. {Neilsen},
\newblock {Physical Review D}, {\bf 80(2)}, 024012 (2009).

\bibitem{ONeill2009ApJ...700..859O}
S.~M. {O'Neill}, M.~C. {Miller}, T. {Bogdanovi{\'c}}, C.~S. {Reynolds}, and J.~D. {Schnittman},
\newblock {Astrophys. J.}, {\bf 700}, 859 (2009).

\bibitem{Corrales2010MNRAS.404..947C}
L.~R. {Corrales}, Z. {Haiman}, and A. {MacFadyen},
\newblock {Monthly Not. Roy. Astron. Soc.}, {\bf 404}, 947 (2010).

\bibitem{Rosotti2012MNRAS.425.1958R}
G.~P. {Rosotti}, G. {Lodato}, and D.~J. {Price},
\newblock {Monthly Not. Roy. Astron. Soc.}, {\bf 425}, 1958 (2012).

\bibitem{Fitchett1983MNRAS.203.1049F}
M.~J. {Fitchett},
\newblock {Monthly Not. Roy. Astron. Soc.}, {\bf 203}, 1049 (1983).

\bibitem{Pietila1995CeMDA..62..377P}
H. {Pietil{\"a}}, P. {Hein{\"a}m{\"a}ki}, S. {Mikkola}, and M.~J. {Valtonen},
\newblock {Celestial Mechanics and Dynamical Astronomy}, {\bf 62}, 377 (1995).

\bibitem{DeMink2017ApJ...839L...7D}
S.~E. {de Mink} and A. {King},
\newblock {Astrophys. J.}, {\bf 839}, 7 (2017).

\bibitem{Lipunov1992ans..book.....L}
V.~M. {Lipunov}, G.~{B{\"o}rner}, and R.~S. {Wadhwa},
\newblock {\em {Astrophysics of Neutron Stars}}
\newblock (Springer, 1992).

\bibitem{Bardeen1970Natur.226...64B}
J.~M. {Bardeen},
\newblock {Nature}, {\bf 226}, 64 (1970).

\bibitem{Thorne1974ApJ...191..507T}
K.~S. {Thorne},
\newblock {Astrophys. J.}, {\bf 191}, 507 (1974).

\bibitem{Li2000ApJ...534L.197L}
L.-X. {Li} and B.~{Paczy{\'n}ski},
\newblock {Astrophys. J. Lett.}, {\bf 534}, L197 (2000).

\bibitem{Bowen2018ApJ...853L..17B}
D.~B. {Bowen}, V.~{Mewes}, M.~{Campanelli}, S.~C. {Noble}, J.~H. {Krolik}, and
  M.~{Zilh{\~a}o},
\newblock {Astrophys. J. Lett.}, {\bf 853}, L17 (2018).

\bibitem{Artymowicz1994ApJ...421..651A}
P.~{Artymowicz} and S.~H. {Lubow},
\newblock {Astrophys. J.}, {\bf 421}, 651 (1994).

\bibitem{Dong2016ApJ...823..141D}
R.~{Dong}, E.~{Vorobyov}, Y.~{Pavlyuchenkov}, E.~{Chiang}, and H.~B. {Liu},
\newblock {Astrophys. J.}, {\bf 823}, 141 (2016).

\bibitem{Armitage2007astro.ph..1485A}
P.~J. {Armitage},
\newblock {e-Print arXiv:astro-ph/0701485} (2007).

\bibitem{Kurbatov2017ARep...61..475K}
E.~P. {Kurbatov} and D.~V. {Bisikalo},
\newblock {Astronomy Reports}, {\bf 61}, 475 (2017).

\bibitem{Samarskii1980MoIzN....W....S}
A.~A. {Samarskii} and I.~P. {Popov},
\newblock {\em {Difference methods for solving problems of gas dynamics /2nd revised and enlarged edition/}}
\newblock (Moscow, Izdatel Nauka, 1980).

\bibitem{Ginzburg1979tpa..book.....G}
V.~L. {Ginzburg},
\newblock {\em {Theoretical physics and astrophysics}}
\newblock (Pergamon, Oxford, 1979).

\bibitem{Illarionov1972SvA....16...45I}
A.~F. {Illarionov} and R.~A. {Syunyaev},
\newblock {Soviet Ast.}, {\bf 16}, 45 (1972).

\bibitem{Pavlov1989PhR...182..187P}
G.~G. {Pavlov}, Y.~A. {Shibanov}, and P.~{M{\'e}sz{\'a}ros},
\newblock {Phys. Rep.}, {\bf 182}, 187 (1989).

\bibitem{Feng2006ApJ...650L..75F}
H.~{Feng} and P.~{Kaaret},
\newblock {Astrophys. J. Lett.}, {\bf 650}, L75 (2006).

\bibitem{Fabrika2017ASPC..510..395F}
S.~{Fabrika},
\newblock {Astronomical Society of the Pacific Conference Series}, {\bf 510}, 395 (2017).

\bibitem{XMM-Newton2018UsersHandbook}
XMM-Newton Community~Support Team,
\newblock {\em {XMM-Newton Users Handbook. Issue 2.16}}
\newblock
  URL: \url{https://xmm-tools.cosmos.esa.int/external/xmm_user_support/documentation/uhb/XMM_UHB.html},
\newblock Accessed: 2018-08-10.

\bibitem{Maraschi1990ApJ...353..452M}
L.~{Maraschi} and S.~{Molendi},
\newblock {Astrophys. J.}, {\bf 353}, 452 (1990).

\bibitem{Starling2011MNRAS.411.2792S}
R.~L.~C. {Starling}, K.~{Wiersema}, A.~J. {Levan}, T.~{Sakamoto}, D.~{Bersier},
  P.~{Goldoni}, S.~R. {Oates}, A.~{Rowlinson}, S.~{Campana}, J.~{Sollerman},
  N.~R. {Tanvir}, D.~{Malesani}, J.~P.~U. {Fynbo}, S.~{Covino}, P.~{D'Avanzo},
  P.~T. {O'Brien}, K.~L. {Page}, J.~P. {Osborne}, S.~D. {Vergani},
  S.~{Barthelmy}, D.~N. {Burrows}, Z.~{Cano}, P.~A. {Curran}, M.~{de Pasquale},
  V.~{D'Elia}, P.~A. {Evans}, H.~{Flores}, A.~S. {Fruchter}, P.~{Garnavich},
  N.~{Gehrels}, J.~{Gorosabel}, J.~{Hjorth}, S.~T. {Holland}, A.~J. {van der
  Horst}, C.~P. {Hurkett}, P.~{Jakobsson}, A.~P. {Kamble}, C.~{Kouveliotou},
  N.~P.~M. {Kuin}, L.~{Kaper}, P.~A. {Mazzali}, P.~E. {Nugent}, E.~{Pian},
  M.~{Stamatikos}, C.~C. {Th{\"o}ne}, and S.~E. {Woosley},
\newblock {Monthly Not. Roy. Astron. Soc.}, {\bf 411}, 2792 (2011).

\bibitem{Ubertini2018IBIS}
P.~{Ubertini},
\newblock {\em {IBIS: Imager on Board the INTEGRAL Satellite}}
\newblock URL: \url{https://www.cosmos.esa.int/web/integral/instruments-ibis},
\newblock Accessed: 2018-08-10.

\bibitem{Ubertini2003A&A...411L.131U}
P.~{Ubertini}, F.~{Lebrun}, G.~{Di Cocco}, A.~{Bazzano}, A.~J. {Bird},
  K.~{Broenstad}, A.~{Goldwurm}, G.~{La Rosa}, C.~{Labanti}, P.~{Laurent},
  I.~F. {Mirabel}, E.~M. {Quadrini}, B.~{Ramsey}, V.~{Reglero}, L.~{Sabau},
  B.~{Sacco}, R.~{Staubert}, L.~{Vigroux}, M.~C. {Weisskopf}, and A.~A.
  {Zdziarski},
\newblock {Astron. Astrophys.}, {\bf 411}, L131 (2003).

\end{thebibliography}
\end{document}